\documentclass{ctr}

\usepackage{ctrfont}
\usepackage{natbib}

\usepackage{url}
\usepackage{amsmath}
\usepackage{amssymb}

\usepackage{graphicx}

\title{On the evolution of the velocity gradient tensor in transitional boundary layers}
\shorttitle{Evolution of the velocity gradient tensor in transitional boundary layers}
\author{A. Elnahhas, P.~L. Johnson, A. Lozano-Dur\'an \and P. Moin}
\shortauthor{Elnahhas et al.}

\begin{document}
\pagenumbering{gobble}


\maketitle

\section{Motivation and objectives} 

The transition of a boundary layer from a laminar to a turbulent state
is associated with a rapid increase in the friction and heat transfer
coefficients due to the enhanced mixing caused by turbulence. This
transition is accompanied by the rapid growth of velocity gradients
throughout the boundary layer, which can be appreciated by visualizing
isosurfaces of the $Q$-criterion or other vortex identifiers and is
particularly evident in late-stage transition, when growing structures
abruptly break down into more chaotic flow, generating turbulent spots
\citep{wu2017pnas}. This drastic change is important from both the
applied and fundamental points of view. In this report, we approach
the transition problem by studying the dynamics of the velocity
gradient tensor.

Canonical transition from laminar to turbulence is conceptually
divided into several steps: the receptivity process, in which the
perturbations are introduced into the boundary layer (for example, by
Tollmien-Schlichting (TS) wave modes), followed by a primary
instability, nonlinear saturation resulting in a new base state,
secondary instability, and eventually breakdown into turbulence
\citep{Schmid2001}.  In the first steps, linear stability analysis
predicts the unstable nature of the laminar boundary layer
profile. Local stability analysis of the linearized Navier-Stokes
equations leads to the emergence of two-dimensional waves, which grow
exponentially in time \citep{MackLM.AGARD1984}. These TS waves
modulate the base flow of the laminar boundary layer and grow to a
finite amplitude, which gives rise to a new base state. The linear
analysis of this new base state is typically framed as a secondary
instability. This results in various types of transition mechanisms,
depending on whether the secondary instability has the same
fundamental frequency as the TS wave or is a subharmonic of that
frequency, respectively called K-type and H-type instabilities
\citep{Herbert1988}.

Most linear stability analyses deal with solutions which are
exponentially growing in either space or time. This exponential growth
continues until nonlinear mechanisms become important and saturate the
amplitude of the unstable modes before the breakdown occurs. Hence,
linear stability analysis does not characterize the saturation
process. In the present brief, we propose to study the velocity
gradient tensor to characterize the late nonlinear stages of
transition.

In the fully turbulent regime, the Lagrangian dynamics of the velocity
gradient tensor have been studied extensively in canonical scenarios
from homogeneous isotropic turbulence (HIT) to turbulent boundary
layers. The governing equations of the velocity gradient tensor are
directly derived from the Navier-Stokes equations. In the seminal
works of \cite{Vieillefosse1982a} and \cite{Cantwell1992} a simplified
version of the equations, in which the nonlocal contributions of
pressure and viscosity are neglected, was solved analytically. This
simplified restricted Euler system reproduces frequently observed
features in direct numerical simulation (DNS) databases of simple
turbulent flows and high-dissipation regions of complex turbulent
flows. These features include the positivity and small value of the
intermediate strain rate eigenvalue, as well as the alignment of the
vorticity vector with the intermediate eigenvector of the strain rate
\citep{Ashurst1987}. The analysis of the evolution of the velocity
gradient tensor can be recast in the form of its invariants, $Q$ and
$R$, which represent the balance between rotation- and strain rate
production-dominated regions, respectively. When the restricted Euler
system is recast in this form, it provides an insightful picture of
the topological states through which a Lagrangian fluid particle could
evolve.

In this preliminary study, we are interested in quantifying the growth
of velocity gradients as a function of the downstream coordinate,
along with the dominant terms that lead to this growth at each stage
of the boundary layer.  It will also be shown that the balance of
enstrophy production and viscous dissipation presented by
\cite{Tennekes1972}, valid at high Reynolds numbers, also holds
immediately after transition. To quantify the streamwise growth of
velocity gradients, several scalar quantities are considered from an
Eulerian budget point of view for a zero-pressure-gradient flat-plate
boundary layer. This analysis is carried out using a DNS data of
H-type transition. It is argued that, based on the modeling experience
of the nonlocal pressure and viscous terms in homogeneous turbulence,
the budget which exhibits the least streamwise pressure dependence
should be pursued further, due to the potential of easier modeling in
subsequent works.

The remainder of this report is structured as follows. Section 2
presents the argument behind why the Frobenius norm of the velocity
gradient tensor is chosen to be analyzed, along with the derivation of
its spanwise/temporally averaged Eulerian streamwise budget. Section 3
describes the numerical framework used to generate the DNS database,
along with the results of computing the budget presented in Section
2. Finally, conclusions are presented in Section 4.


\section{Eulerian streamwise velocity gradient tensor budget equations}

Consider the incompressible Navier-Stokes equations,
\begin{equation}
    \frac{\partial u_i}{\partial t}+u_j\frac{\partial u_i}{\partial x_j}=-\frac{\partial p}{\partial x_i}+\nu\frac{\partial^2 u_i}{\partial x_j \partial x_j},
\end{equation}
where $u_i$ is the velocity component in the $i$th direction, $x,y,
\mathrm{and}~z$ are the streamwise, wall-normal, and spanwise
directions, respectively, $\nu$ is the kinematic viscosity, and $p$ is
the kinematic pressure. The frame of reference adopted in the
subsequent analysis is the one fixed to the origin of the boundary
layer. Taking the gradient of Eq. (2.1) leads to a governing equation
for the velocity gradient tensor, 
\begin{equation}
  A_{ij}=\frac{\partial u_i}{\partial x_j}.
\end{equation}
The equation for its evolution is given by
\begin{equation}
  \frac{\partial A_{ij}}{\partial
    t}+u_k\frac{\partial A_{ij}}{\partial
    x_k}=-A_{ik}A_{kj}-\frac{1}{\rho}\frac{\partial^2 p}{\partial x_i
    \partial x_j}+\nu\frac{\partial^2 A_{ij}}{\partial x_k \partial
    x_k}\mathrm{,}
\end{equation}
where we follow the nomenclature by \cite{Meneveau2011}. From a
Lagrangian perspective, Eq. (2.3) is not closed in terms of $A_{ij}$
at position $\vec{x}$ and time $t$ due to the last two terms on the
right-hand side, those being the pressure Hessian and the viscous
terms. Taking the trace of Eq. (2.3) while accounting for the
incompressibility condition $A_{ii} = 0$ leads to the Poisson equation
for pressure
\begin{equation} \frac{\partial^2 p}{\partial x_l
    \partial x_l}=-A_{lk}A_{kl},
\end{equation}
which can be used to separate the local and nonlocal parts, in a
Lagrangian manner, of Eq. (2.3) as follows
\begin{equation}
  \frac{\partial A_{ij}}{\partial t}+u_k\frac{\partial
    A_{ij}}{\partial
    x_k}=-(A_{ik}A_{kj}-\frac{1}{3}A_{mk}A_{km}\delta_{ij})+H_{ij}^p+H_{ij}^\nu,
\end{equation}
where
\begin{equation}
    H_{ij}^p=-\bigg(\frac{\partial^2 p}{\partial x_i \partial x_j}-\frac{1}{3}\frac{\partial^2 p}{\partial x_k \partial x_k}\delta_{ij}\bigg) \;\; \mathrm{and} \;\; H_{ij}^\nu=\nu\frac{\partial^2 A_{ij}}{\partial x_k \partial x_k},
\end{equation}
are the anisotropic part of the pressure Hessian and the viscous term,
respectively.

In this report, we are interested in characterizing the growth of the
velocity gradients in an aggregate manner as a function of the
downstream coordinate. We aim at identifying a scalar quantity which
to characterize nonlinear stages of transition. To this end, it is
possible to derive the evolution equation of several scalar quantities
by making appropriate contractions between various terms and
Eq. (2.5). The resulting equations are then averaged along the
homogeneous directions and in time, then integrated across the
boundary layer thickness to establish the Eulerian streamwise budgets.

It is beneficial for these quantities, along with their governing
equations, to possess certain properties. First, the statistical mean
of the quantity being examined needs to be appreciably different in
the laminar and turbulent regions of the flow. Second, the terms
contributing to the growth or decay of this quantity should be
physically interpretable to elucidate the processes at place. Third,
and perhaps with a more practical outlook in mind, the dominant terms
responsible for the growth and decay of the selected quantity should
be amenable to modeling. The following subsections discuss possible
candidates for quantities to be used as transition markers.

\subsection{The $Q$ equation}
Contracting Eq. (2.5) with the transpose of the velocity gradient
tensor $A_{ji}$ yields the evolution equation for the second invariant
of the velocity gradient tensor
\begin{equation}
    \frac{\partial Q}{\partial t}+u_k\frac{\partial Q}{\partial x_k}=-3R-A_{ji}H_{ij}^p-A_{ji}H_{ij}^\nu\mathrm{,}
\end{equation}
where
\begin{equation}
    Q=-\frac{1}{2}A_{im}A_{mi}=\frac{1}{2}(\Omega_{im}\Omega_{im}-S_{im}S_{im}) \;\; \mathrm{and} \;\; R=-\frac{1}{3}A_{im}A_{mn}A_{ni}\mathrm{,}
\end{equation}
are the second and third invariants of the velocity gradient
tensor. $\Omega_{ij}$ and $S_{ij}$ are the rotation and strain rate
tensors, respectively. Regions of high $Q$ are rotation dominated, and
regions of high $R$ are strain production dominated.

From the perspective of the Eulerian budgets discussed above, $Q$ is
not a good indicator for transition. Consider the case of
statistically stationary, homogeneous turbulence. $\overline{Q}$,
where the overbar denotes averaging in the homogeneous directions and
in time, is a function of only the mean velocity gradients, and as
such, its expectation is zero for most homogeneous flows, indicating
that the contribution of rotation- and strain- dominated regions to an
integral measure based on $\overline{Q}$ will balance. This
approximately holds at high Reynolds numbers in a turbulent boundary
layer. In the laminar region of a boundary layer, $\overline{Q}$ is of
the same order of magnitude as that found in the turbulent region;
therefore, there is no differentiation between the two cases when
considering the evolution of an integral quantity based on
$Q$. Furthermore, the viscous term in Eq. (2.7) can be rewritten to
have a term akin to dissipation in the kinetic energy
equation
\begin{equation} \nu\frac{\partial A_{ij}}{\partial
    x_k}\frac{\partial A_{ji}}{\partial x_k},
\end{equation}
which cannot be associated with a particular sign and is, therefore,
lacking physical interpretability. Thus, an equation based on $Q$ does
not satisfy two of the three properties outlined above. In order to
remedy these two deficiencies, we consider the evolution equations for
$Q^2$ and the Frobenius norm of the velocity gradient tensor,
$|A|_F^2=A_{ij}A_{ij}$.

\subsection{The $Q^2$ equation}

It has been argued that looking at an integral quantity based on
$\overline{Q}$ in a boundary layer would not provide a contrasting
view between the laminar and turbulent portions. In contrast, it is
apparent that rotating and straining regions are ubiquitous in the
turbulent boundary layer compared to the laminar one. As such, an
integral measure that sums up the contributions of both rotation- and
strain-dominated regions could differentiate between the laminar and
turbulent regions. Therefore, let us consider the evolution equation
for $Q^2$. By contracting Eq. (2.7) with $Q$, we get
\begin{equation}
  \frac{\partial Q^2}{\partial t}+u_k\frac{\partial Q^2}{\partial x_k}
  = -6QR-2Q(A_{ji}H_{ij}^p+A_{ji}H_{ij}^\nu),
\end{equation}
which possesses a positive source term in the mean because one can see
that $Q$ and $R$ are negatively correlated by looking at their
characteristic teardrop-shaped joint probability distribution function
\citep{Soria1994}.

Even though considering $\overline{Q^2}$ alleviates the problem
associated with $\overline{Q}$ having a similar value in the laminar
and turbulent portions of a boundary layer, a problem with the
pressure term arises. The pressure term in Eq. (2.10), after the
streamwise budget equation for $\overline{Q^2}$ is formulated,
involves a correlation between $Q$ and the divergence of the velocity
and pressure gradients. Through numerically analyzing the
$\overline{Q^2}$ budget, it is found that this correlation is a
dominant term. Recalling that the anisotropic portion of the pressure
Hessian leads to a closure problem from a Lagrangian perspective, many
attempts have been made at modeling its effect. In HIT, examples of
these models include the tetrad model by \cite{Chertkov1999} and the
fluid deformation approximation by \cite{Chevillard2006}. However,
unlike the viscous term, which acts to damp all the trajectories in
the $Q$ and $R$ phase space by driving them toward the origin, the
effect of the deviatoric part of the pressure Hessian differs
depending on the region of the $Q$ and $R$ phase space being examined
\citep{Chevillard2008}. This difficulty associated with modeling the
pressure Hessian in HIT is expected to be exacerbated when transition
in a boundary layer is considered. Since we are examining a
zero-pressure-gradient flat-plate boundary layer, and keeping these
future modeling difficulties in mind, a governing equation that allows
for the neglect of the pressure term is pursued.

\subsection{The $|A|_F^2$ equation}

Contracting Eq. (2.5) with the velocity gradient tensor $A_{ij}$
itself leads to
\begin{equation}
    \frac{\partial |A|_F^2}{\partial t}+u_k\frac{\partial |A|_F^2}{\partial x_k}=-2A_{ik}A_{kj}A_{ij}+2A_{ij}H_{ij}^p+2A_{ij}H_{ij}^\nu,
\end{equation}
where $|A|_F^2=A_{ij}A_{ij}$. The pressure Hessian and the viscous
terms in Eq. (2.11) can be written in the following form
\begin{equation}
    \frac{\partial |A|_F^2}{\partial t}+u_k\frac{\partial |A|_F^2}{\partial u_k}=-2A_{ik}A_{kj}A_{ij}-2\frac{\partial}{\partial x_i}\bigg(A_{ij}\frac{\partial p}{\partial x_j}\bigg)+\nu\frac{\partial^2 |A|_F^2}{\partial x_k \partial x_k}-2\nu\frac{\partial A_{ij}}{\partial x_k}\frac{\partial A_{ij}}{\partial x_k},
\end{equation}
which displays the favorable properties sought above. 

First, $A_{ij}A_{ij}=S_{ij}S_{ij}+\Omega_{ij}\Omega_{ij}$, indicating
that an integral measure based on the Frobenius norm of the velocity
gradient tensor adds up the effects of both strain- and
rotation-dominated regions, differentiating between the laminar and
turbulent portions of a boundary layer. Second, unlike the viscous
term in the $Q$ equation, Eq. (2.9), the viscous dissipation term in
Eq. (2.12) is associated with a negative sign and can, therefore, be
interpreted clearly. Third, the pressure term in Eq. (2.12) is in
divergence form, meaning that it can be neglected in the case of a
zero-pressure-gradient flat-plate boundary layer due to the nominal
scaling of pressure gradients in boundary layers. Finally, it is
important to discuss the source term of Eq. (2.12),
$A_{ik}A_{kj}A_{ij}$, and its physical meaning. $A_{ik}A_{kj}A_{ij}$
can be written as
\begin{equation}
  A_{ik}A_{kj}A_{ij}=S_{ik}S_{kj}S_{ji}-\Omega_{ik}S_{kj}\Omega_{ji}=S_{ik}S_{kj}S_{ij}-\frac{1}{4}\omega_kS_{kj}\omega_j\mathrm{,}
\end{equation}
where $\omega_i$ is the vorticity in the $i$th direction. Equation
(2.13) shows that the source term in the governing equation for
$|A|_F^2$ can be decomposed into two contributions. First,
$S_{ik}S_{kj}S_{ij}$ is the strain self-amplification term which
contributes to the production of strain and can be shown to be
negative on average by examining the term in its principal axes frame
and noting that \cite{Ashurst1987} observed the positivity of the
intermediate eigenvalue of the strain rate tensor. Second,
$\omega_kS_{kj}\omega_j$ is the vortex stretching term leading to
enstrophy production, which is on average positive due to both the
preferential alignment of the vorticity vector with the intermediate
eigenvector of the strain rate tensor and the axial vortex stretching
due to the primary strain rate eigenvector, the latter being the
dominant mechanism \citep{Doan2018}. Thus, on average, the source term
of Eq. (2.12) is a positive quantity.

To proceed from Eq. (2.12), to an integral equation based on
$\overline{|A|^2_F}$, we take the mean in the homogeneous spanwise
direction and in time, noting that the boundary layer is statistically
stationary. Then we integrate in the wall-normal direction across the
boundary layer. This leads to the following equation
\begin{equation}
\begin{gathered}
            \frac{d}{dx}\bigg(\int_0^\infty\overline{u|A|^2_F}dy\bigg)=-2\int_0^\infty\overline{A_{ik}A_{kj}A_{ij}}dy-2\frac{d}{dx}\bigg(\int_0^\infty\overline{\frac{\partial u}{\partial x_k}\frac{\partial p}{\partial x_k}}dy\bigg) \\
            -\nu\frac{\partial \overline{|A|^2_F}}{\partial y}\bigg|_0+\nu\frac{d^2}{dx^2}\bigg(\int_0^\infty\overline{|A|^2_F}dy\bigg)-2\nu\int_0^\infty\overline{\frac{\partial A_{ij}}{\partial x_k}\frac{\partial A_{ij}}{\partial x_k}}dy,
\end{gathered}
\end{equation}
where both the pressure gradient term and the streamwise viscous term
are kept for thoroughness, although we know that they are negligible
on the basis of nominal boundary layer scaling. We will refer to these
terms as, going from left to right, streamwise growth, gradient
self-amplification, streamwise pressure transport, wall viscous
deposition, streamwise viscous transport, and viscous destruction of
velocity gradients.

The left-hand side of this equation can be thought of as the rate of
change of the velocity gradient tensor magnitude as we march
downstream along the boundary layer, in the frame of reference
attached to the origin of the boundary layer. Thus, there is no time
dependency due to statistical stationarity. As such, even though the
left-hand side of Eq. (2.14) is not Galilean invariant, it is only so
because an average in time was taken, and in non-stationary flows, the
addition of the time derivative would retrieve Galilean invariance. By
observing the rapid generation of velocity gradients in a boundary
layer around the transitional region, it is expected that the
left-hand side term is small in both the laminar and fully turbulent
regions of the boundary layer when compared to its magnitude during
transition. In the fully turbulent region in the limit of high
Reynolds numbers, the scaling presented by \cite{Tennekes1972} about
the balance of enstrophy production and viscous dissipation suggests
that the gradient self-amplification term and the viscous destruction
term should be the dominant terms that balance in the
$\overline{|A|^2_F}$ budget. By computing each of the terms on the
right-hand side, we aim to identify the important mechanisms that
drive transition through the generation of velocity gradients, as well
as determine how far upstream this high Reynolds number scaling holds.

\section{Numerical simulation}

The numerical experiment follows a similar setup as the one presented
by \cite{Lozano-Duran2018a}. In particular, it follows the case
presented in Figure 1(b) of that paper, where the parabolized
stability equations are used to march the initial boundary condition
up to a matching location from which a DNS is carried out. The
governing equations are integrated with a staggered second-order
central finite-difference method, and time advancement utilized a
second-order Runge-Kutta scheme combined with the fractional-step
procedure. A zero-pressure-gradient flat-plate boundary layer is
simulated as it undergoes H-type laminar-to-turbulent transition
\citep{Herbert1988}. Velocities are nondimensionalized using the
free-stream velocity $U_\infty$, and wall units are denoted by a
superscript + and defined in terms of the friction velocity $u_\tau$
and kinematic viscosity $\nu$. Transition is triggered by imposing an
inflow boundary condition that is the sum of the Blasius solution, a
fundamental TS wave of nondimensional frequency
$2F=\omega\nu/U_\infty^2$ and subharmonic mode of frequency $F$. These
modes are the solution to the local Orr-Sommerfeld-Squire problem at
$Re_x=1.8\times10^5$.

Similarly to \cite{Lozano-Duran2018a}, the simulation box is periodic
in the spanwise direction. It is narrower in the spanwise extent than
in the cases presented by \cite{Lozano-Duran2018a}, and its size is
equal to the inverse of the oblique wavenumber $\beta$ associated with
the subharmonic mode. Thus, there exists only a single
$\Lambda$-vortex at any streamwise location in the simulation
domain. Table 1 shows some of the important computational parameters,
such as domain size, where $\delta_{in}$ indicates the initial
boundary layer thickness at the Reynolds number corresponding to the
matching location between PSE and DNS, $Re_{x,match}$.

\begin{table}
    \centering
    \begin{tabular}{c|c}
        $\Delta y^+_{min}$ & $0.35$ \\ 
        $\Delta x^+$ & $7.0$ \\
        $\Delta z^+$ & $7.6$ \\
        $L_x$ & $142\delta_{in}$ \\
        $L_y$ & $18\delta_{in}$ \\
        $L_z$ & $6.5\delta_{in}$ \\
        $\delta_{out}/\delta_{in}$ & $\approx3$ \\
        $2F$ & $1.2395\times10^{-4}$ \\
        $\beta L_z/(2\pi)$ & $1.0078$ \\
        $Re_{x,match}$ & $3.9\times10^5$
    \end{tabular}
    \caption{Characteristic parameters of the simulation}
    \label{tab:parameteres}
\end{table}
\begin{figure}
    \centering
    \includegraphics[width = \textwidth]{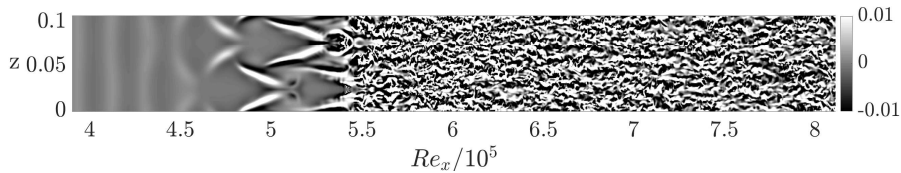}
    \caption{Wall-normal velocity at $y=0.02 \delta_{in}$.}
    \label{fig:flow field}
\end{figure}

Figure 1 shows a snapshot of the wall-normal velocity of the flow
field, illustrating the staggered arrangement of $\Lambda$-vortices
associated with H-type transition. Figure 2 shows the skin friction
coefficient as a function of $Re_x$, along with the laminar and
turbulent correlations, as well as the integral streamwise flux of the
velocity gradients across the boundary layer as a function of
$Re_x$. Even though the simulation domain is minimal in the spanwise
extent, in the sense that we are enforcing the periodicity of each
$\Lambda$-vortex, the skin friction coefficient displays the main
features of controlled natural transition in larger domains, such as
the skin friction overshoot \citep{Sayadi2013}. Furthermore, the
profile of the integral streamwise flux of the velocity gradients
across the boundary resembles the skin friction profile, indicating
the rapid flux of velocity gradients around the transitional
region. However, unlike the skin friction coefficient, this streamwise
flux is not dependent on gradients at the wall.

\begin{figure}
    \centering
    \includegraphics[width = \textwidth]{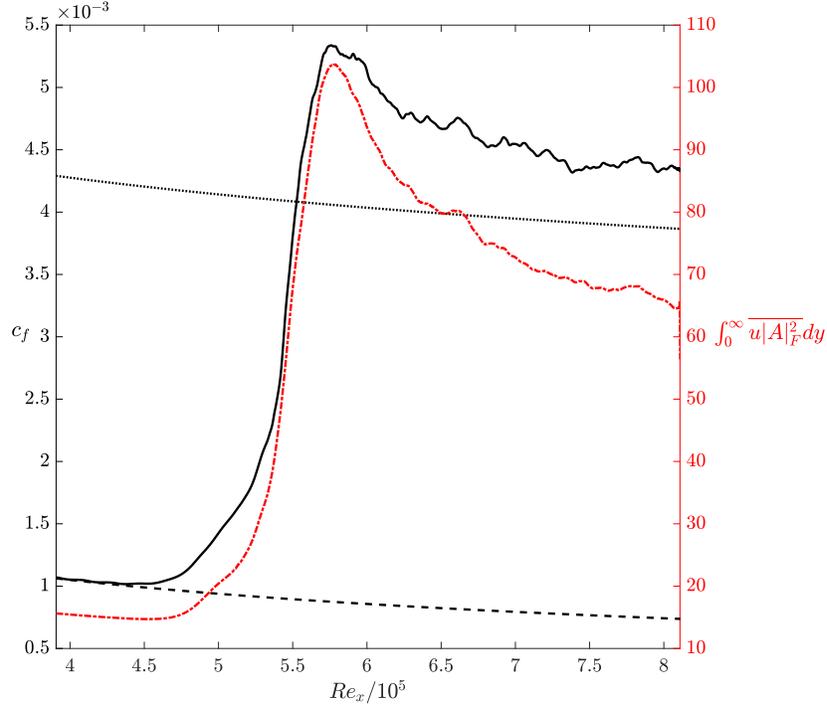}
    \caption{Left $y$-axis: Skin friction coefficient versus $Re_x$. ---, H-type; ..., Blasius; - -, turbulent correlation \citep{White1974}.
    Right $y$-axis: -.-, Integral streamwise flux of velocity gradients across the boundary layer versus $Re_x$.}
    \label{fig:skin friction}
\end{figure}

\subsection{The $|A|^2_F$ budget}

To compute the terms in Eq. (2.14), 200 evenly spaced time instances
spanning four periods of the TS wave were used. Figure 3 shows these
terms as a function of $Re_x$. Figure 4 shows the same budget
normalized by the gradient self-amplification as a function of $Re_x$
and is truncated to start at $Re_x = 4.5\times10^5$ because that is
the Reynolds number corresponding to the position when the gradient
self-amplification term starts to become large. We can infer multiple
things from Figures 3 and 4. First, consider the integral
measure. Compared to its value around transition, its values in the
laminar and turbulent regions are negligible, which matches our prior
expectation and indicates that the largest changes in velocity
gradients in a boundary are around the transition point. Second,
consider the magnitudes of the gradient self-amplification and viscous
destruction terms as compared to all the other terms in the budget. It
is expected that at very high Reynolds numbers, these two terms should
dominate the budget and match due to the same scaling arguments used
to balance enstrophy production and viscous dissipation
\citep{Tennekes1972}. This high Reynolds number scaling is apparent
from Figure 4. However, it also extends back to the late stages of
transition around $Re_x\approx5.5\times10^5$, corresponding to the
breaking of the $\Lambda$-vortices in Figure 1. This indicates that
even at the late stages of transition, it is the same kinematic
effects of vortex stretching and strain self-amplification that
dominate the generation of large velocity gradients in the boundary
layer.

This aforementioned result opens two avenues for further
exploration. First, the study of models based on the Lagrangian
dynamics of the velocity gradient tensor and their applicability in
the transitional region could be explored, perhaps in the context of
localized breakdown events. Second, the dynamics of the late stages of
transition can be studied to formulate reduced-order dynamical models
that are applicable at the fully developed turbulent regions of the
boundary layer \citep{Sayadi2013}. This second approach could be
pursued from the perspective of coherent structures and has been
explored by \cite{Sayadi2014} using dynamic mode decomposition.

It is important to note that only qualitative results were inferred
from Figures 3 and 4, due to the two sides of Eq. (2.14) not balancing
numerically. However, in the region of interest
$Re_x\geqslant4.5\times10^5$, the relative magnitude of the difference
between the left- and right-hand sides of the equation is
approximately $8\%$ when compared to the largest term in the budget,
the gradient self-amplification term. This difference is expected for
several reasons. First, Eq. (2.14) holds in a continuous sense, and as
such, we do not expect the two sides of the equation to balance unless
it was derived discretely. Second, Eq. (2.14) is valid for a
statistically stationary flow, and the minor oscillations still
present in the budget indicates that further averaging is required to
warrant neglecting the time derivative on the left-hand side. Third,
and more importantly, the terms present in Eq. (2.14) include
higher-order derivatives, such as the viscous destruction term, that
require high-order numerical schemes, as well as higher resolution, to
be accurately captured \citep{Lozano-Duran2015,
  Lozano-Duran2016}. Taking these limitations into account, the
qualitative statements made can still be justified.

\begin{figure}
    \centering
    \includegraphics[width = \textwidth]{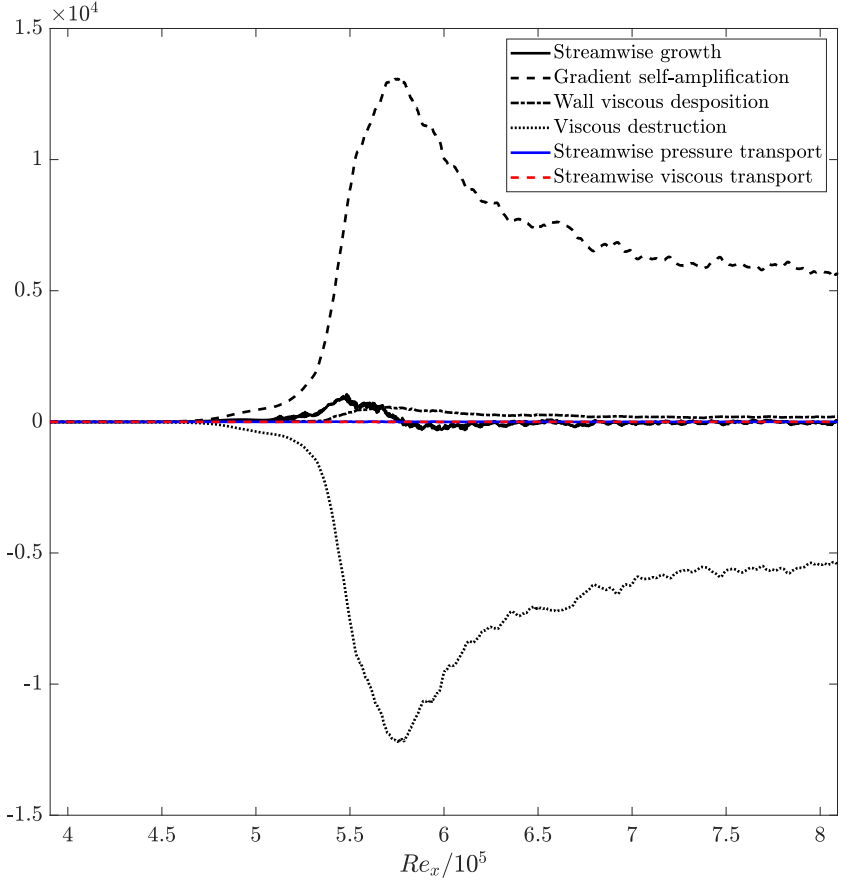}
    \caption{Eulerian budget of the Frobenius norm of the velocity gradient tensor, Eq. (2.14), versus $Re_x$.}
    \label{fig:budget}
\end{figure}

\begin{figure}
    \centering
    \includegraphics[width = \textwidth]{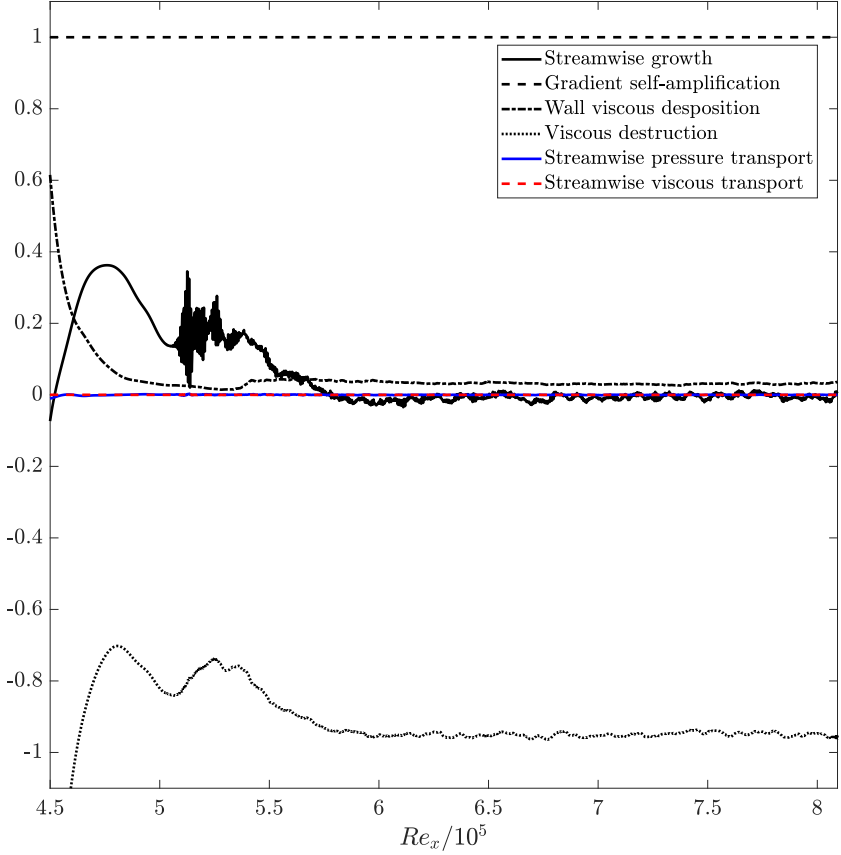}
    \caption{Eulerian budget of the Frobenius norm of the velocity gradient tensor, Eq. (2.14), normalized by the gradient self-amplification term versus $Re_x$.}
    \label{fig:my_label}
\end{figure}

\section{Conclusions}

In this report, we study the transition to turbulence from the
perspective of the velocity gradient tensor dynamics. Our work is
motivated by the observation of nonlinear structures emerging during
transition, as revealed by vortex identifiers such as the
$Q$-criterion. To that end, we have derived transport equations based
on several invariants of the velocity gradient tensor to obtain
integral budgets spanning the different stages of transition from
laminar to turbulent flow. We have also discussed which quantity would
be the most appropriate in our study, while keeping in mind the
potential for future modeling applications.

The Frobenius norm of the velocity gradient tensor, $|A|_F^2$, was
identified as a suitable marker for transition due to the independence
of its growth from the pressure term. It was found that the source
term for $|A|_F^2$, which was a combination of the vorticity
stretching and strain self-amplification mechanisms, is balanced by
the viscous destruction/dissipation term, and that the two terms are
at least one order of magnitude larger than the remaining terms in the
integral budget. This result, already documented at the high Reynolds
number turbulence, was also found to hold at the late stages of
transition associated with the initial breakup of the
$\Lambda$-vortices. The outcome suggests that some of the dynamics
that govern fully developed turbulence are also present in the late
stages of transition, and that these late-stage dynamics might aid the
development of reduced-order models for fully developed turbulence. At
the same time, it provides support to the study of transition from the
point of view of Lagrangian velocity gradient tensor dynamics.

\section*{Acknowledgments}
P.L.J. was funded by the Advanced Simulation and Computing (ASC)
program of the US Department of Energy's National Nuclear Security
Administration via the PSAAP-II Center at Stanford, Grant
No.~DE-NA0002373. A.L.-D. acknowledges the support of NASA under grant
No.~NNX15AU93A and of ONR under grant No.~N00014-16-S-BA10.


\bibliographystyle{ctr}

\end{document}